\documentclass[11pt]{article}
\usepackage{amsmath,amsfonts,amssymb}
\usepackage{theorem}
\usepackage{url}
\usepackage{verbatim}
\usepackage[a4paper]{geometry}
\usepackage{hyperref}

\newcommand{\FF}{\mathbb {F}}
\newcommand{\N}{\mathbb {N}}

\newcommand{\F}{\mathcal {F}}
\newcommand{\C}{\mathcal {C}}
\newcommand{\M}{\mathcal {M}}
\newcommand{\ddiv}{\operatorname {div}}
\newcommand{\Norm}{\operatorname {N}}
\newcommand{\Res}{\operatorname {Res}}

\newtheorem{theorem}{Theorem}{\bfseries}{\itshape}
\newtheorem{heuristic}[theorem]{Heuristic}{\bfseries}{\itshape}
\newtheorem{prop}[theorem]{Proposition}{\bfseries}{\itshape}
\newtheorem{cor}[theorem]{Corollary}{\bfseries}{\itshape}
\def\qed{\hfill{$\Box$}\bgroup\parfillskip=0pt\par\egroup}
\newenvironment{proof}[1][Proof]{\noindent{\em #1}. }{\qed\smallskip\par}
\newenvironment {remark}
{\paragraph {Remark.}}{}

\makeatletter
\let\@fnsymbol\@arabic
\makeatother

\title{An $L (1/3)$ Discrete Logarithm Algorithm for Low Degree Curves}

\author {%
Andreas Enge\footnote{INRIA / CNRS / Universit\'e de Bordeaux}\ ,
Pierrick Gaudry\footnote{INRIA / CNRS / Nancy Universit\'e}\ \ and
Emmanuel Thom{\'e}\footnotemark[2]%
}

\date {December 13, 2009}

\begin{document}
\maketitle

\begin{abstract}
We present an algorithm for solving the discrete logarithm problem in
Jacobians of families of plane curves whose degrees in $X$ and $Y$ are
low with respect to their genera. The finite base fields $\FF_q$ are
arbitrary, but their sizes should not grow too fast compared to the
genus.  For such families, the group structure and discrete logarithms can
be computed in subexponential time of $L_{q^g}(1/3, O(1))$. The runtime
bounds rely on heuristics similar to the ones used in the number field
sieve or the function field sieve.
\end{abstract}

\section {Introduction}

The discrete logarithm problem (DLP) is the keystone for the security of
cryptosystems based on elliptic curves and on Jacobian groups of more
general algebraic curves.  While to date, elliptic curves provide a very
broad range of groups for which no algorithm improves over the generic
ones for attacking the DLP, the same does not hold for higher genus
curves. A variety of algorithms exists to tackle the DLP on Jacobians of
curves, depending on whether the problem is being considered with the
field size or the genus growing to infinity, or possibly both.
For a general overview on algorithms for the DLP, see the survey
\cite{Enge08}. The
outcome is that for implementing cryptographic primitives, curves
of genus~3 and higher have clear practical
disadvantages over curves of genus~2 and elliptic curves. Yet, studying
the DLP on these curves is important in particular because of the Weil
descent strategy, which reduces the DLP on elliptic curves over extension
fields to the DLP in the Jacobian of a curve of higher genus. Therefore,
besides the better understanding of the general picture that one may
obtain by studying large genus curves, an algorithm for solving the DLP in
the large genus case may eventually become a threat for some
elliptic curve cryptosystems.

The following is a general strategy for solving the DLP in groups
enjoying in particular a suitable notion of size (for more
details on an appropriate model, see \cite{EnGa02}). A
first phase consists in collecting relations involving elements of a
chosen factor base, which is a subset of the group under
consideration formed by elements of relatively small size.
Thereafter, the logarithms of these elements are deduced by linear
algebra.  Depending on the exact algorithm employed, the output of this
computation either gives the logarithm of a chosen set of group elements,
or in more advanced algorithms, the ability to compute the logarithms of
arbitrary elements at a relatively low cost. The resulting complexity
is usually of subexponential nature, namely of the form
\[
L_N (\alpha, c) = e^{c (\log N)^\alpha (\log \log N)^{1 - \alpha}}
\]
for $\alpha \in (0, 1)$ and $c > 0$, where $N$ is the
group size.

Quite early on, it appeared that this approach could be
adapted to a family of hyperelliptic curves over a fixed base field
$\FF_q$ and of genus $g$ growing to infinity.  In this case the
algorithm from~\cite{AdDeHu94} solves the DLP in subexponential time
$L_{q^g}(1/2,O(1))$.  This complexity is heuristic. It is established
under the assumption that a given family of polynomials behaves similarly
to random polynomials of the same degree. Later on, rigorous results for
smoothness of divisors have led to proofs of the subexponential running
time, and the algorithm has been generalised to further classes of
curves~\cite{EnSt02,MuStTh99,Enge02,EnGa02,Couveignes01,Hess04}.
These results imply that given a family of algebraic curves of growing genus
$g$ over a base field $\FF_q$ with $\log q$ bounded by some polynomial in
$g$, solving the DLP is possible in proven subexponential time
$L_{q^g}(1/2,O(1))$.

We briefly mention, at the opposite end of the spectrum, the DLP on a
family of curves of fixed genus over a base field $\FF_q$ with $q$
growing to infinity. In this case, analogous algorithms have a complexity
which is exponential in $\log
q$ \cite{GaThThDi07,Diem06,DiTh08}. This case is not studied here.

Subexponential algorithms are known in other common contexts, namely integer
factorisation and computation of discrete logarithms in finite fields.
Proven algorithms of complexity $L(1/2)$ exist, however the most
efficient algorithms for these problems are the number field
sieve~\cite{BuLePo93,Gordon93b} and the function field
sieve~\cite{AdHu99} and their derivatives, which achieve a heuristic
complexity of $L(1/3)$. For a long time, it has been an
open problem to decide whether such a complexity can be
achieved for solving the discrete logarithm problem in Jacobian groups
of algebraic curves.

We answer this question positively for a relatively large class of
curves and present a probabilistic algorithm of heuristic
subexponential complexity $L_{q^g}(1/3,O(1))$ for solving the discrete
logarithm problem in Jacobians of curves of genus $g$ over finite fields
$\FF_q$. Here, we consider families of curves $\C_i(X,Y)$ of genus
$g_i$ over finite fields $\FF_{q_i}$. We require $g_i\geq(\log q_i)^2$,
and the degrees in $X$ and $Y$ must stay within the non-empty interval
with end points $\approx g_i^\alpha$ and $\approx g_i^{1 - \alpha}$,
where
$1/3 \leq \alpha \leq 2/3$. Our constraint on the curve equation is the
key for producing principal divisors of small degree, in a
manner analogous to the function field sieve. The computation of
individual logarithms, once the relation collection and linear algebra
steps have been completed, is performed using a special-$Q$
descent strategy.

A previous related result appeared in~\cite{EnGa07}; however, this
earlier version has been considerably improved. First, the class of
curves to which our algorithm applies has been expanded. Furthermore the
computation of discrete logarithms no longer has complexity
$L(1/3+\varepsilon,o(1))$, but rather $L(1/3,O(1))$. This raises the question
of determining explicitly the constant represented by $O(1)$.
Assuming the family of
curves satisfies $\deg_X\C_i\cdot\deg_Y\C_i\leq\kappa g_i$, the exact
complexity of our algorithm is $L(1/3,(64\kappa/9)^{1/3})$, which is a
familiar complexity in the context of the number field sieve.  We mention
that subsequently to~\cite{EnGa07}, Diem has
presented at the 10th Workshop on Elliptic Curve Cryptography (ECC 2006)
an algorithm based on similar ideas~\cite{Diem06ecc};
he argued that computing discrete logarithms for non-singular plane
curves can be solved in $L(1/3,(64/9)^{1/3}+\varepsilon)$ for any
$\varepsilon>0$. We show that the same complexity is also
achieved using a slight modification of our algorithm
and that it is valid for a
class of curves strictly including those handled by Diem's algorithm.

The article is organised as follows. Section~\ref{sec:idea} gives an
informal presentation of the algorithm. Section~\ref{sec:smoothness}
provides the necessary tools for the precise statement and analysis
of the algorithm, which is given in Sections~\ref{sec:rel}
and~\ref{sec:logarithms}. Some corner and special cases are
studied in Section~\ref{sec:critical}.

\section {Main idea}

\label{sec:idea}

\subsection{Relation collection}
\label {ssec:idea-relation}

Before describing our algorithm with all its technical details on the most
general class of curves, we sketch in this section the main idea yielding a
complexity of $L_{q^g}(1/3, O(1))$ for a restricted class of curves.
We provide a simplified analysis by
hand waving; Section~\ref {sec:smoothness} is devoted to a more precise
description of the heuristics used and of the smoothness properties needed
for the analysis.

Let $\FF_q$ be a fixed finite field. We consider a family of $C_{nd}$ curves
over $\FF_q$, that is, curves of the form
\[
\C : Y^n + X^d + f (X, Y)
\]
without affine singularities such that
$\gcd (n, d) = 1$ and any monomial $X^i Y^j$ occurring in $f$ satisfies
$n i + d j < nd$ (see~\cite{Miura98}).
Such a curve has genus $g = \frac {(n-1)(d-1)}{2}$; we assume that $g$
tends to infinity, and that $n \approx g^{\alpha}$ and $d \approx
g^{1-\alpha}$ for some $\alpha\in \left[ \frac13,\frac23 \right]$
(we use the symbol $\approx$, meaning ``about the same size''
with no precise definition).
The non-singular model of a $C_{nd}$ curve has a unique point at infinity,
which is $\FF_q$-rational; so there is a natural bijection between degree
zero divisors and affine divisors, and in the following, we shall only be
concerned with effective affine divisors. Choose as factor base $\F$ the
about $L_{q^g}(1/3, O(1))$ prime divisors of smallest degree, that is, of
degree bounded by some $B \in \N$ with
$B \approx \log_q L_{q^g}(1/3, O(1))$.

To obtain relations, consider functions $\varphi(X,Y)\in\FF_q[X,Y]$ such that
\[
k=\deg_Y\varphi\approx g^{\alpha-1/3}\quad\text{and}\quad
\delta=\deg_X\varphi\approx g^{2/3-\alpha}.\]
Whenever the affine part $\ddiv (\varphi)$ of the divisor of $\varphi$ is
smooth with respect to the factor base, it yields a relation, and we have
to estimate the probability of this event.

Let $\Norm$ be the norm of the function field extension $\FF_q (\C) =
\FF_q (X)[Y] / (Y^n + X^d + f (X, Y))$ relative to $\FF_q (X)$.
For a given function $\varphi$ on the curve, if $\ddiv\varphi$ contains
only places of inertia degree~$1$, then $\ddiv\varphi$ is $B$-smooth if
and only if the norm of $\varphi$ is.  We have
\begin {eqnarray*}
\deg_X \Norm (\varphi) & = & \deg \Res_Y (\varphi(X,Y), Y^n + X^d + f (X,
Y)), \\
& \leq & \deg_X\varphi\deg_Y\C + \deg_Y\varphi\deg_X\C
= n \delta + kd
\approx g^{2/3}.
\end {eqnarray*}
Heuristically, we assume that the norm behaves like a random polynomial of
degree about $g^{2/3}$. Then it is $B$-smooth with probability
$1 / L_{q^g}(1/3, O(1))$. (This is the same theorem as the one stating that a random
polynomial of degree $g$ is $\log_q L_{q^g}(1/2, O(1))$-smooth with probability
$1 / L_{q^g}(1/2, O(1))$, cf., for instance, Theorem~2.1 of \cite {BePo98}.)
Equivalently, we may assume heuristically that $\ddiv (\varphi)$ behaves like a
random effective divisor of the same degree $\deg_X \Norm (\varphi)$.
Then the standard results on
arithmetic semigroups (cf. Section~\ref {sec:smoothness}) yield again that
$\ddiv (\varphi)$ is smooth with probability $1 / L_{q^g}(1/3, O(1))$.

So the expected time for obtaining $|\F| = L_{q^g}(1/3, O(1))$
relations is $L_{q^g}(1/3, O(1))$. With the same complexity, one can
solve a linear system and obtain the discrete logarithms of
the elements of $\F$. If the group structure was not known in advance, it
is also possible to deduce it from a Smith normal form computation, which
lies again in the same complexity class.

It remains to show that the search space is sufficiently large to yield the
required $L_{q^g}(1/3, O(1))$ relations, or otherwise said, that the number of
candidates for $\varphi$ is at least $L_{q^g}(1/3, O(1))$. The number of
$\varphi$ is about
$$
q^{k \delta} \approx q^{g^{1/3}} <  e^{ (g^{1/3} (\log q)^{1/3}) (\log (g \log q))^{2/3}}
= L_{q^g}(1/3, O(1)).
$$
The previous inequality in the place of the desired equality shows that a
more rigorous analysis requires a careful handling of the $\log q$
factors in the exponent; in particular, $k$ or $\delta$ has to be slightly
increased. Moreover,
the constant exponent in the subexponential function needs to be taken into
account.

\subsection{Individual logarithms}
\label {ssec:idea-descent}

After Section~\ref {ssec:idea-relation}, the discrete logarithms of the
elements of the factor base $\F$ are known.  Now, to solve a
general discrete logarithm problem, we need to be able to rewrite any
element in terms of elements of $\F$. The classical tool for doing so is the
special-$Q$ descent strategy as introduced by
Coppersmith~\cite{Coppersmith84}.

The input is a place $Q = \ddiv(u(X), Y-v(X))$, for which the
discrete logarithm is sought. While not all elements can be written in that
form, most of them can; so without loss of generality, by randomising
the input, we may assume the special form. The degree of $Q$ is $\deg u \leq g$, and $\deg v < \deg u$.

One step of the special-$Q$ descent rewrites a place of degree
$\approx g^{1/3+\tau}$ for some $\tau \in [0, 2/3]$ as a sum of places of
degrees bounded by $g^{1/3+\tau/2}$.
Thus, the place $Q$ of degree at most $g$ is first rewritten as a sum of
places of degrees bounded by $g^{2/3}$. Each of them is then rewritten as a
sum of places of degrees bounded by $g^{1/2}$, and so on. We end up
with a tree of places, whose leaves have a degree as close to $g^{1/3}$ as
we wish. Therefore, pushing the special-$Q$ descent far enough,
we can hope to obtain leaves that are in $\F$, so that the discrete
logarithms of all the elements of the tree, including that of $Q$, can be
deduced.

Let us now sketch how one step of the special-$Q$ descent works in our
case: We consider a place $Q = \ddiv(u(X), Y-v(X))$, with $\deg v < \deg
u \approx g^{1/3+\tau}$ for some $\tau \in [0,2/3]$. The polynomial
functions on the curve having a zero at $Q$ and of degree in $Y$ bounded
by $k\approx g^{\alpha-1/3+\tau/2}$ form an $\FF_q[X]$-lattice generated
by
$$\left(u(X), Y-v(X),
Y^2- \big( v(X)^2 \bmod u(X) \big), \ldots,
Y^k - \big( v(X)^k \bmod u(X) \big)\right).$$
We consider $\FF_q[X]$-linear combinations of these basis elements that
have a small degree in $X$: Allowing coefficients in the combination to
have a degree up to $\approx g^{2/3-\alpha+\tau/2}$, the
corresponding functions have a degree in $X$ bounded by $\approx
g^{2/3-\alpha+\tau/2}$.
Among the $\approx q^{g^{1/3 + \tau}}$ such functions, we limit
ourselves to a sieving space of size about $q^{g^{1/3}}$.

The degree of the affine part of the divisor of each function $\varphi$
in the sieving space is bounded by $n\deg_X\varphi +
d\deg_Y\varphi\approx g^{2/3+\tau/2}$. Since there are about $q^{g^{1/3}}$
of them, one can expect to find one whose divisor is $\approx
g^{1/3+\tau/2}$-smooth (apart from the place $Q$ that is present in the
divisor by construction). We have then rewritten $Q$ as a sum of divisors
of degree at most $\approx g^{1/3+\tau/2}$ in time $L(1/3)$.

In this description, we have been vague with respect to the degree bounds,
and it is necessary to be more accurate, especially when $\tau$ is
getting close to $0$.  This motivates the following section, in which we
examine in more detail the smoothness results and heuristics that are
needed for the algorithm.

\section {Smoothness}
\label {sec:smoothness}

The algorithm presented in this article relies on finding relations as
smooth divisors of random polynomial functions of low degree. As with
other algorithms of this kind, for instance \cite{AdDeHu94}, its running
time analysis will be heuristic. The main heuristic assumption
is that certain principal
divisors are as likely to be smooth as random divisors of the same
degree, for which the desired smoothness probabilities can be proved.
In this section we collect the needed smoothness results, before
discussing our heuristics in more detail.

We suppose
that all curves are given by absolutely irreducible plane affine models
\[
\C : F (X, Y)
\]
with $F \in \FF_q [X, Y]$, where $\FF_q$ is the exact constant field of
the function field of $\C$. Arithmetic of elements of the Jacobian group
of such curves is detailed in~\cite{Hess02}. In particular, operations
such as splitting a divisor into a sum of places can be performed in
polynomial time.

Essentially, we are interested in a factor base $\F$ consisting of the
places of degree bounded by some parameter $\mu$ (a few technical
modifications are necessary and will be discussed later in this section).
Then an effective divisor of degree $\nu$ is called $\F$-smooth or
$\mu$-smooth if it is composed only of places in $\F$.
The probability of $\mu$-smoothness is ruled by the usual results on
smoothness probabilities in arithmetic semigroups such as the integers or
polynomials over a finite field, cf. \cite {Manstavicius92}.

Unfortunately, most results in the literature are stated for a fixed
semigroup and give asymptotics for $\mu$ and $\nu$ tending to infinity,
whereas we need information that is uniform over an infinite family of
curves. Notice, however, the purely combinatorial nature of the question:
How many objects of size up to $\nu$ can be built from irreducible blocks
of size up to $\mu$? The answer depends only on the number of building blocks
of any given size, and it turns out that its main term is the same uniformly
over all semigroups under consideration. This can be exploited to prove
combinatorially,
in the same spirit as for hyperelliptic curves in \cite{EnSt02}, the following
general result, which is Theorem~13 of~\cite{Hess04}:

\begin {theorem}[He\ss]
\label {th:hess}
Let $0 < \varepsilon < 1$, $\gamma = \frac {3}{1 - \varepsilon}$ and $\nu$,
$\mu$ and $u = \frac {\nu}{\mu}$ such that
$3 \log_q (14 g + 4) \leq \mu \leq \nu^\varepsilon$ and
$u \geq 2 \log (g + 1)$.
Denote by $\psi (\nu, \mu)$ the number of $\mu$-smooth effective divisors of
degree $\nu$. Then for $\mu$ and $\nu$ sufficiently large (with an explicit
bound depending only on $\varepsilon$, but not on $q$ or $g$),
\[
\frac {\psi (\nu, \mu)}{q^\nu} \geq e^{- u \log u \left( 1 +
\frac {\log \log u + \gamma}{\log u} \right)}
= e^{- u \log u (1 + o (1))}.
\]
\end {theorem}

Denote by
\[
L (\alpha, c) = L_{q^g} (\alpha, c)
= e^{c (g \log q)^\alpha (\log (g \log q))^{1 - \alpha}}
\]
for $0 \leq \alpha \leq 1$ and $c > 0$
the subexponential function with respect to $g \log q$, and
let
\[
\M = \M_{q^g} = \log_q (g \log q) = \frac {\log (g \log q)}{\log q}.
\]
The parameter $g \log q$ will be the input size for the class of curves
we consider; more intrinsically, this is
the logarithmic size
of the group in which the discrete logarithm problem is defined.

\begin {prop}
\label {prop:smoothness}
For some $0 < \beta < \alpha \leq 1$ and $c$, $d > 0$, let
\[
\nu = \lfloor \log_q L (\alpha, c) \rfloor
    = \lfloor c g^\alpha \M^{1 - \alpha} \rfloor
\text { and }
\mu = \lceil \log_q L (\beta, d) \rceil
    = \lceil d g^\beta \M^{1 - \beta} \rceil.
\]
Assume that there is a constant
$\rho > \frac {1 - \alpha}{\alpha - \beta}$ such that
$g \geq (\log q)^\rho$.
Then for $g$ sufficiently large,
\[
\frac {\psi (\nu, \mu)}{q^\nu} \geq L \left( \alpha - \beta, - \frac {c}{d}
(\alpha - \beta) + o (1) \right),
\]
where $o (1)$ is a function that is bounded in absolute value by a constant
(depending on $\alpha$, $\beta$, $c$, $d$ and $\rho$) times
$\frac {\log \log (g \log q)}{\log (g \log q)}$.
\end {prop}

\begin {proof}
One computes
\[
u = \frac {\nu}{\mu}
  \leq \frac {c}{d} \left( \frac {g \log q}{\log (g \log q)}
  \right)^{\alpha - \beta}
\]
(the inequality being due only to the rounding of $\nu$ and $\mu$),
\[
\log u = (\alpha - \beta) \log (g \log q) (1 + o (1))
\]
and
\[
\frac {\log \log u}{\log u} = o (1),
\]
with both $o (1)$ terms being of the form stipulated in the proposition.
Applying Theorem~\ref {th:hess} yields the desired result. Its
prerequisites are satisfied since
\begin {eqnarray*}
\limsup
\frac {\log \mu}{\log \nu}
& = & \limsup
\frac {\beta \log g - (1 - \beta) \log
  \log q}{\alpha \log g - (1 - \alpha) \log \log q} \\
& \leq & \limsup
\frac {\beta \log g}{\alpha \log g -
\frac {1 - \alpha}{\rho} \log g} \\
& = & \frac {\beta}{\alpha - \frac {1 - \alpha}{\rho}}
=: \varepsilon' < 1
\end {eqnarray*}
because of the definition of $\rho$; then $\varepsilon$ is taken to
be any value strictly larger than $\varepsilon'$ and less than 1.
\end {proof}

The choice of $\mu$ shall insure that the factor base size, that is about
$q^\mu$, becomes subexponential. But the necessary rounding of $\mu$, which
may increase $q^\mu$ by a factor of almost $q$, may result in more than
subexponentially many elements in the factor base when $q$ grows too fast
compared to~$g$.

\begin {prop}
\label {prop:subexponentiality}
Let $0 < \beta < 1$ and $\rho \geq \frac {1 - \beta}{\beta}$.
If $g \geq (\log q)^\rho$, then
$q = L (\beta, o (1))$ for $g \to \infty$.
\end {prop}

\begin {proof}
One computes
$$ q =  e^{\log q} = e^{(\log q)^{1 - \beta} (\log q)^\beta}.$$
Since $g\geq (\log q)^\rho$ with $\rho\geq\frac{1-\beta}{\beta}$, one
gets $(\log q)^{1-\beta}\leq g^\beta$, so that
$q \leq e^{(g\log q)^\beta}$. Compared to $L(\beta, 1)$, the term
$(\log(g\log q))^{1-\beta}$ is missing in the exponent; since this term
tends to infinity, the result follows.
\end {proof}

\begin {cor}
\label {cor:rounding}
Let $0 < \beta < 1$,
$\rho > \frac {1 - \alpha}{\alpha - \beta}$ and
$\rho \geq \frac {1 - \beta}{\beta}$,
and $g \geq (\log q)^\rho$.
Then Proposition~\ref {prop:smoothness} remains valid for
an arbitrary rounding of $\mu$ and $\nu$, and
$q^\mu = L (\beta, d + o (1))$.
\end {cor}

\begin {proof}
Let $k$ be any integer. By Proposition~\ref {prop:subexponentiality},
\[
\nu + k = \left\lfloor \log_q (q^k L (\alpha, c)) \right\rfloor
= \left\lfloor \log_q L (\alpha, c + o (1)) \right\rfloor,
\]
which shows that $\nu$ may be replaced by $\nu + k$ in
Proposition~\ref {prop:smoothness}.
The same argumentation holds for $\mu$.
\end {proof}

We need to deal with a few technicalities related to the potential
singularities and the places at infinity of our curves. To this
purpose, we augment the factor base as follows; this addition of
a polynomial number of divisors is negligible compared to the
subexponential factor base size. Furthermore, the computational expense
incurred by these additions is also negligible, since the algorithms
in~\cite{Hess02} have polynomial complexity.
\begin {itemize}
\item
Add to $\F$ all the places corresponding to the resolution of
singularities, regardless of
their degrees, whose number is bounded by $\frac {(d-1)(d-2)}{2}$ with $d =
\deg F$. The algorithm can then be described as if the curves
were non-singular.
\item
Add to $\F$ the infinite places corresponding to non-singularities, regardless
of their degrees, whose number is bounded by $d$ by B{\'e}zout's theorem.
Then a divisor is $\F$-smooth if and only if its affine part is.
\end {itemize}

The correctness and the running time analysis of our algorithm depend
on two heuristics, that are classical in the context of discrete
logarithm computations by collecting smooth relations.
First of all, the smoothness probabilities of
Proposition~\ref {prop:smoothness} should also apply to the special
way in which we create the relations.

\begin {heuristic}
\label {heu:smoothness}
Let $D$ of degree $\nu$ be the affine part of the divisor of a uniformly
randomly chosen polynomial $\varphi$ with imposed bounds on the degrees
in $X$ and $Y$. Then the probability of $D$ to be $\F$-smooth is
asymptotically the same as that of a random effective affine divisor of
degree $\nu$ to be $\mu$-smooth.  If $\varphi$ is additionally
constrained to have a zero in a special place $Q$, the same holds for
$\ddiv \varphi - Q$.
\end {heuristic}

The first part of the heuristic covers the initial relation collection
phase as described in Section~\ref {ssec:idea-relation}, the second
part is needed for the special $Q$-descent of
Section~\ref {ssec:idea-descent} for computing individual logarithms.
They ensure that relations are found sufficiently quickly.
Next, one needs to make sure that the found relations are
sufficiently varied to capture the complete Jacobian group.

\begin {heuristic}
\label {heu:independence}
The probability that the relations found by the algorithm span
the full relation lattice is the same as for random relations.
\end {heuristic}

Here, the \textit {full relation lattice} designates the lattice
$L$ such that the Jacobian group of $\C$ over $\FF_q$ is isomorphic
to the quotient by $L$ of the free abelian group over the factor base.
\textit {Randomness of relations} is to be understood as the uniform
distribution on the set of relations with coefficients between~$0$
and the order of the Jacobian group.

Depending on the choice of $\F$, it is not immediately clear why
Heuristic~\ref {heu:independence} should hold. For instance, assume
that $\F$ contains places of inertia degree larger than~$1$ with
respect to the function field extension $\FF_q (X)[Y] / (\C)$ over
$\FF_q (X)$, that is, places corresponding to ideals $(u, v (X, Y))$
with $u \in \FF_q [X]$ and $\deg_Y v > 1$. If $\varphi$ is limited
to being linear in~$Y$, then no such place may occur in a relation,
so that the relation lattice cannot have full rank.

In practice, however, inert places should be very rare. This is justified
by the observation that these places have a Dirichlet density of~$0$:
A place of degree $\mu$ and inertia degree $f$ dividing $\mu$
corresponds to a closed point on $\C$ with $X$-coordinate in $\FF_{q^{\mu/f}}$
and $Y$-coordinate in $\FF_{q^\mu}$, of which there are on the order of
$q^{\mu/f}$. Clearly, places with $f \geq 2$ are completely negligible.
Notice now that the proof of Theorem~\ref {th:hess} is entirely combinatorial
and relies on the fact that there are essentially $q^\mu / \mu$ places of
degree $\mu$. As this is still the case when restricting to non-inert places,
the proof of the theorem should carry over. This motivates an
\textit {a priori} artificial
restriction of the factor base to non-inert places.

To summarise, we rely on the validity of Heuristics~\ref {heu:smoothness}
and~\ref {heu:independence} for the factor base $\F$ of smoothness
parameter $\mu$ containing the following places:
\begin {itemize}
\item
all places corresponding to the resolution of singularities;
\item
all places at infinity (i.e., places where the function $X$ has a negative
valuation).
\item
the affine non-inert places of degree bounded by $\mu$, or otherwise said,
the places corresponding to prime ideals of the form $(u, Y - v)$ with
$u \in \FF_q [X]$ irreducible of degree at most $\mu$ and
$v \in \FF_q [X]$ of degree less than $\deg u$.
\end {itemize}

\section {Relation search}

For the time being, we assume that all groups we are dealing with are
cyclic, of known order and with a known generator which is part of the
factor base. Discrete logarithms are taken with respect to this generator.
We discuss the complications arising when one of these conditions is not
satisfied at the end of Section~\ref {sec:logarithms}.

We are now ready to formulate precisely the algorithm outlined
in Section~\ref{sec:idea}, together
with its complexity analysis. We start by the relation collection and
linear algebra phases as sketched in Section~\ref{ssec:idea-relation}.

\label{sec:rel}
\begin {theorem}
\label {th:relations}
Let $(\C_i (X, Y))_{i \in \N}$ be a family of plane curves of genus
$g_i$ over $\FF_{q_i}$ of degrees $n_i$ in $Y$ and $d_i$ in $X$.
Assume that there are constants $\kappa > 0$ and $\rho \geq 2$ such that
\begin {eqnarray}
\label {eq:kappa}
&& \frac {n_i d_i}{g_i} \leq \kappa \\
\label {eq:limit}
&& \frac {n_i}{(g_i / \M_i)^{1/3}} \to \infty,
\frac {d_i}{(g_i / \M_i)^{1/3}} \to \infty
\text { with } \M_i = \frac {\log (g_i \log q_i)}{\log q_i} \\
\label {eq:mu}
&& g_i \geq (\log q_i)^\rho
\end {eqnarray}
Let $b$ be defined by
$$b =  \sqrt [3]{\frac {8 \kappa }{9}}.$$
There exists an algorithm that computes a factor base with $L_{q_i^{g_i}}
(1/3, b)$ elements, together with the discrete logarithms of all
the factor base elements, in an expected running time of
$$
L_{q_i^{g_i}} (1/3, c + o (1))
\text { with }
c = \sqrt [3]{\frac {64\kappa}{9}}
$$
under Heuristics~\ref{heu:smoothness} and~\ref{heu:independence}.
\end {theorem}

\begin {proof}
For the sake of notational clarity, we drop all indices $i$ in the
following.

Let $\nu$, $\delta > 0$ be constants to be optimised later. Consider
polynomials $\varphi (X, Y) \in \FF_q [X, Y]$, seen as functions on $\C$,
of degrees bounded by
$\left\lceil \nu \frac {n}{(g/\M)^{1/3}} \right\rceil$ in $Y$ and
$\left\lceil \delta \frac {\kappa g/n}{(g/\M)^{1/3}} \right\rceil$ in
$X$.
Then~(\ref{eq:limit}) implies that
\begin {equation}
\label {eq:critical}
\deg_Y \varphi \leq \nu \frac {n}{(g/\M)^{1/3}} (1 + o (1))
\text { and }
\deg_X \varphi
\leq \delta \frac {\kappa g/n}{(g/\M)^{1/3}} (1 + o (1)).
\end {equation}
The affine part of the divisor of $\varphi$ has a degree bounded by
\begin {align*}
\deg_X \Res_Y (\varphi, \C)
& \leq  \deg_X \varphi \deg_Y \C + \deg_Y \varphi \deg_X \C \\
& \leq \left(\delta \kappa g^{2/3} \M^{1/3} + \nu n d g^{-1/3}
\M^{1/3}\right)\cdot(1+o(1)) \\
& \leq  \kappa (\delta + \nu + o(1)) g^{2/3} \M^{1/3} \text { by (\ref {eq:kappa})} \\
& =  \log_q L (2/3, \kappa (\delta + \nu +o(1))).
\end {align*}

Let $b > 0$ be a constant to be optimised later, and choose a smoothness
bound of $\lceil \log_q (L (1/3, b)) \rceil$. Then by \eqref {eq:mu} and
Corollary~\ref {cor:rounding}, the factor base size is in $L (1/3, b + o (1))$,
and by Corollary~\ref {cor:rounding} and Heuristic~\ref {heu:smoothness},
the smoothness probability of the divisor of $\varphi$ is at least
\[
L \left( 1/3, - \frac {\kappa (\nu + \delta)}{3 b}+o(1)\right).
\]
The number of different $\varphi$ that satisfy the chosen degree
bounds is at least
\[
q^{\kappa \nu \delta g^{1/3} \M^{2/3}}
= L (1/3, \kappa \nu \delta).
\]
So the expected number of relations obtained from all these $\varphi$ is
bounded below by
$L \left( 1/3, \kappa \left( \nu \delta - \frac {\nu + \delta}{3 b}
\right)+o(1)\right)$.
For the linear algebra to succeed, according to
Heuristic~\ref {heu:independence}, we need the number of relations
to exceed the factor base size. To
minimise the relation collection effort, we choose $\nu$ and
$\delta$ such that equality holds, that is,
\begin {equation}
\label {eq:minsieve}
\kappa \nu \delta - \frac {\kappa (\nu + \delta)}{3 b} = b.
\end {equation}
On the other hand, we wish to choose the parameters such that
the time taken by the (sparse) linear algebra phase, which is $L (1/3,
2b+o(1))$, is comparable with the time taken by the relation collection:
\begin {equation}
\label {eq:sieve=linalg}
\kappa \nu \delta = 2 b.
\end {equation}

Substituting $\kappa \nu \delta$ from (\ref {eq:sieve=linalg})
into (\ref {eq:minsieve}), we obtain
\[
\nu + \delta = \frac {3 b^2}{\kappa}.
\]

So the sum and product of
$\nu$ and $\delta$ are known, and $\nu$ and $\delta$ are the
roots of the quadratic polynomial
\[
X^2 - \frac {3 b^2}{\kappa} X + \frac {2 b}{\kappa}.
\]
For the roots to exist as real numbers, the discriminant of the
quadratic polynomial must be non-negative, which is equivalent to
\[
b \geq \sqrt [3]{\frac {8 \kappa}{9}}.
\]
Since we want to minimise the effort, we choose $b$ minimal and reach
equality above.
Then
\[
\nu = \delta = \sqrt {\frac {2 b}{\kappa}}
= \sqrt [3]{\frac {8}{3 \kappa}}.
\]
The total running time becomes $L (1/3, c + o (1))$ with
\[
c = 2 b = \sqrt [3]{\frac {64 \kappa }{9 }}.
\]
\end {proof}

\section{Computing discrete logarithms}
\label {sec:dlog}

We now turn to the precise description and analysis of the special-$Q$
descent strategy outlined in Section~\ref{ssec:idea-descent}.

\label{sec:logarithms}
\begin {theorem}
\label {th:logarithms}
Under the assumptions of Theorem~\ref{th:relations}, once the
relation collection and linear algebra steps have been
completed, the logarithm of any divisor in the Jacobian group of $\C_i$
over $\FF_{q_i}$ can be computed in time
\[
L_{q_i^{g_i}} (1/3, b + \varepsilon)
\text { with }
b = \sqrt [3]{\frac {8 \kappa }{9}}
\text { and any }
\varepsilon > 0.
\]
\end {theorem}
Notice that this complexity is well below that of
Theorem~\ref {th:relations} for the relation collection and linear algebra
phases.

\begin{proof}
Without loss of generality, one may
assume that the element whose logarithm is sought is a place
of degree bounded by $g$ and of inertia degree~$1$,
cf. the discussion at the end of Section~\ref {sec:smoothness}.

More precisely, let $Q=\ddiv(u(X), Y-v(X))$ be a place with
$\deg v < \deg u \leq \log_q L(1/3+\tau, c)$ for some
$c>0$ and $0\le\tau\le 2/3$. The place we start with has
$\tau=\frac23$ and $c=1$.

We consider the polynomial functions on the curve having a zero at $Q$,
and in particular the lattice of polynomials $\varphi$
of degree in $Y$ bounded by $k$ with
$$ k = \left\lfloor \sigma \frac{n}{(g/\M)^{1/3-\tau/2}} \right\rfloor,$$
where $\sigma > 0$ is a constant to be determined later.
These $\varphi$ form an $\FF_q[X]$-lattice generated by
$\left(v_0(X), Y-v_1(X), Y^2-v_2 (X), \ldots, Y^k - v_k(X) \right)$
with $v_0 = u$ and $v_i = v^i \bmod u$ for $i \geq 1$.

Let $L(1/3,e + o (1))$ be the effort we are willing to expend for one smoothing
step, where $e > 0$ is a parameter to be optimised later.
Then we need a sieving space of the same size, and are thus
looking for $L(1/3,e + o (1))$
distinct $(k+1)$-tuples of polynomials $\left(\alpha_0(X), \alpha_1(X),
\ldots, \alpha_k(X)\right)$ and corresponding functions
\[
\varphi = -
\alpha_0(X)v_0(x)+\sum_{i=1}^k \alpha_i(X) (Y^i - v_i(X))
= \sum_{i=1}^k \alpha_i (X) Y^i - \sum_{i=0}^k \alpha_i (X) v_i (X).
\]
At the same time, we wish to minimise the degree of $\varphi$ in $X$.
Recall that the degree of $v_i$ is bounded by
$D := \log_q L(1/3+\tau, c)$. Then for any integer $z$, linear algebra
on the lattice yields $q^{kz}$ different tuples such that the degrees
of the $\alpha_i$ and that of $\sum_i \alpha_i v_i$ are at most
$\frac {D}{k} + z$.
Choose $z$ so as to obtain a sieving space of size $L(1/3,e + o (1))$,
that is, solve $q^{kz}=L(1/3, e + o (1))$, or
\[
z =\frac{1}{n}\log_q L(2/3-\tau/2,e/\sigma + o (1)).
\]
Now the degree of $\varphi$ in $X$ is bounded from above by
$\frac {D}{k} + z$ with
$\frac {D}{k} = \frac {1}{n} \log_q L (2/3 + \tau / 2, c / \sigma)$.
Whenever $\tau$ is bounded away from zero, the value of $z$ is thus
negligible compared to that of $D/k$. However, to encompass in a unified
treatment the case where $\tau$ approaches zero, we crudely bound
$-\tau/2$ by $+\tau/2$ in the expression
for $z$ to obtain
$$ \deg_X \varphi \leq
\frac1n{\log_q L(2/3+\tau/2, (c+e)/\sigma + o (1))}.$$

The degree of the affine part of the divisor of $\varphi$ is again, as
in the proof of Theorem~\ref {th:relations}, bounded by
\begin{align*}
\deg_X\varphi\deg_Y\C + \deg_Y\varphi\deg_X\C
&\leq n\deg_X\varphi + kd,\\
&\leq \log_q L(2/3+\tau/2, (c+e)/\sigma + \sigma\kappa + o (1))
\end{align*}
since
$$k d \leq \sigma \frac{nd}{(g/\M)^{1/3-\tau/2}}
\stackrel {\eqref {eq:kappa}}{\leq} \sigma \frac{\kappa g}{(g/\M)^{1/3-\tau/2}}
=\log_q L(2/3+\tau/2, \sigma\kappa).$$

So out of the $L(1/3, e + o (1))$ possible $\varphi$, we expect by
Corollary~\ref {cor:rounding} and Heuristic~\ref {heu:smoothness}
that one is
$\log_q L(1/3+\tau/2, c')$-smooth for
\[
c' = \frac{1}{3e}\left((c+e)/\sigma + \sigma\kappa\right).
\]
To minimise this quantity, we let $\sigma = \sqrt{(c+e)/\kappa}$, so
that
\begin {equation}
\label {eq:cprime}
c' = \frac{2\sqrt{\kappa}}{3e}\sqrt{c+e}.
\end {equation}

Let us summarise the procedure:
Starting with $Q$ of degree $g = \log_q L(1/3+2/3, 1)$,
we use the technique above (with $\tau_0=2/3$, $c_0=1$) to smooth it into
places of degree at most $\log_q L(1/3+\tau_1, c_1)$ with $\tau_1 = 1/3$
and $c_1 = {2\sqrt{\kappa(c_0+e)}}/{3e}$. Each of these is then smoothed again
into places of degree at most $\log_q L(1/3+\tau_2, c_2)$, and so on,
following the formulae
$$ \tau_i = \frac{1}{3\cdot 2^{i-1}},\quad
 c_i = \frac{2\sqrt{\kappa}}{3e}\sqrt{c_{i-1}+e}.$$
After $i$ steps, we get places of degree at most
$$ \log_q L_{q^g}\left(\frac{1}{3} + \frac{1}{3\cdot 2^{i-1}}, c_i\right) =
\log_q L_{q^g}\left(\frac{1}{3}, c_i \,
\M^{\frac{1}{3\cdot 2^{i-1}}}
\right).$$

We need to bound the $c_i$. Studying the function $f(x) =
\alpha\sqrt{x+\beta}$ yields that the sequence $(c_i)$ converges to a
finite limit $c_\infty$, obtained by solving $c'=c$ in \eqref {eq:cprime},
so that
$$ c_\infty = \chi/2\left(\chi +
\sqrt{\chi^2+4e}\right), \quad \text{where}\quad \chi =
\frac{2\sqrt{\kappa}}{3e}.$$ Fix an arbitrary constant $\xi>0$. After
a certain number of steps, depending only on $e$, $\kappa$ and $\xi$, we have
$c_i<c_\infty\cdot(1+\xi)$.  Furthermore, after $O(\log\log g)$ steps, we
can also bound the expression
$\M^{\frac{1}{3\cdot 2^{i-1}}}$ by $(1+\xi)$.

It follows that for any positive constant $\xi$, by building a special-$Q$
descent tree of depth $O(\log\log g)$, we can smooth elements down to a
degree $$ \log_q L_{q^g}\left(\frac{1}{3}, c_\infty(1+\xi)\right).$$

Each node in the tree has arity bounded by $g$, so the number of nodes in
the tree is in $g^{O(\log \log g)}=L_{q^g}(1/3, o(1))$ and
has no influence on the overall complexity.
We finally compute the effort needed to reach $c_\infty=b$. We have
$9b^3=8\kappa$, and we write $9e^3=E\kappa$,
with $E$ to be determined.
The equation $b=c_\infty$ simplifies as:
$$\left(\frac{8}E\right)^{1/3}=\frac2E(1+\sqrt{1+E}).$$
The latter holds for $E=8$, which gives $e=b$.
We therefore conclude that the special-$Q$ descent finishes within time
$L_{q^g}\left(1/3,b+\varepsilon\right)$ for any fixed
$\varepsilon>0$.
\smallskip

So far, we have remained silent about the exact nature of the $o(1)$
terms. As long as a fixed number of them is involved, this does not pose any
problem. But the number of smoothing steps and thus ultimately the number of
applications of Theorem~\ref{th:hess} is not constant. So at first sight,
it is not clear whether the sum of all the $o(1)$ terms is still in
$o(1)$. However, since the depth of the tree is in $O(\log\log g)$, and
since according to Proposition~\ref{prop:smoothness} the $o(1)$ is
actually a constant times  $\frac{\log\log(g\log q)}{\log(g\log q)}$,
the overall function still tends to $0$ and is a $o(1)$.
\end{proof}

\paragraph {The non-cyclic case.}
In general, the Jacobian group need not be cyclic, but may have
up to $2g$ invariant factors. In this case, we call
``discrete logarithm'' of an element its
coefficient vector with respect to a basis of the invariant factor
decomposition. Otherwise said, we need to compute a tuple of scalars
instead of a single one.

We assume that the group order is still known and start by considering
the comparatively easy case that we are given two elements $P$ and $Q$, where
$Q$ is
a multiple of $P$, and we wish to compute the unknown multiplier,
the discrete logarithm of $Q$ to the base $P$.
Write down the relation matrix exactly as in
Theorem~\ref{th:relations}, and perform two descents as in
Theorem~\ref{th:logarithms} for decomposing $P$ and $Q$ as sums of factor
base elements. The right hand sides of the two decompositions are appended
to the relation matrix.
An element of the kernel of this matrix modulo the group
order gives the sought relationship between $P$ and $Q$. The discrete
logarithm can be deduced from it if the coefficient corresponding to
$Q$ is coprime to the group order;
using techniques as in~\cite{EnGa02}, this can be guaranteed to
happen with probability approaching~$1$. The final
complexity is then the same as in Theorem~\ref{th:relations}.

This approach generalises immediately to the non-cyclic case if an
explicit basis $\{ P_i \}$ of the invariant factors is known together
with the exact orders of the basis elements. Then the discrete logarithm
of an element $Q$ as a tuple with respect to the $P_i$ may be obtained
as follows.
After decomposing the $P_i$ and $Q$ over the factor
base as in Theorem~\ref{th:logarithms}, the matrix may be augmented by
the right hand sides of all these decompositions.
An element of the kernel yields the sought expression of $Q$ in terms of
the $P_i$ as long as the coefficient corresponding to $Q$ is coprime with
the group order. Again, the total complexity is as in
Theorem~\ref{th:relations}.

We finally show how to obtain the group structure if only the group order is
known. The classical approach is to compute a Smith Normal Form (SNF)
of the relation matrix obtained in Theorem~\ref{th:relations}, but this is
more costly than a sparse kernel computation. Using the knowledge
of the group order and the fact that for divisor class groups of curves
there is a known set of generators of polynomial size,
He{\ss} shows in \cite[Lemma~50]{Hess04} how to tweak the SNF
computation to keep the same low complexity as before.
In our context, after having computed the relation matrix as in
Theorem~\ref{th:relations} and a set of generators of polynomial
cardinality $r$, we
apply $r$ times Theorem~\ref{th:logarithms} to obtain a decomposition of
each generator in terms of the factor base elements. The right hand sides
of these decompositions are appended to the matrix. Then some
$r$ kernel elements are computed by sparse linear algebra modulo the
group order, yielding
relations between the generators. Using the randomisation techniques
of \cite{EnGa02}, one may ensure that these relations are uniformly
distributed over all kernel elements. It is
then easy to compute a Smith Normal Form (SNF) of this matrix of
polynomial size, thus giving an explicit basis for the group structure.
The overall complexity is then again the same as for
Theorem~\ref{th:relations}.

\paragraph {Group order.}
If the group order is unknown, it may be obtained alongside the
invariant factors from the SNF of the relation matrix of
Theorem~\ref{th:relations}; but computing the SNF, while still being
of complexity $L (1/3)$, would needlessly increase the constant
of the subexponential function.

Instead, one may use the point counting algorithm due to Lauder and
Wan~\cite{LaWa08}, which has a complexity that is polynomial in $p$,
the degree of the finite field extension and the degree of the curve
equation. Notice that by \eqref {eq:kappa}, the latter is in $O (g)$.
If $p$ is very small compared to $g$, for instance, in the
extreme case that $p$ is fixed, then Lauder and Wan's
algorithm has an overall polynomial time complexity. But even in the
most general setting in which Theorem~\ref{th:relations} applies, we
have $q = L (1/3, o (1))$ by Corollary~\ref {cor:rounding}, so that
computing the group order takes only time $L (1/3, o (1))$.

In practice, SNF computations may still be faster than Lauder and Wan's
algorithm in corner cases.
It may then be worthwhile to switch to the algorithm of~\cite{CaHuVe08}
for $\C_{ab}$ curves, which has a quasi-linear complexity in $p$;
or to that of~\cite{Minzlaff08} for superelliptic curves, which has a
square-root complexity in $p$.

\section{Limit cases and special classes of curves}
\label{sec:critical}

\subsection{$n$ close to $(g / \M)^{1/3}$}

In this and the following section, we examine what happens when
the hypothesis \eqref {eq:limit} of Theorem~\ref {th:relations}
is not satisfied. First, we consider the case
$0 < \liminf \frac {n_i}{\left(g_i/\M_i\right)^{1/3}} =: \lambda<\infty$
(the symmetric condition for $d_i$ is handled analogously).
To simplify the presentation, we assume that we have switched to
a subsequence that approaches the limit, and drop again all indices $i$.

Following the proof of Theorem~\ref{th:relations}, we see that the degree
in $Y$ of $\varphi$ poses problem: It tends to $\lceil \nu \lambda \rceil$,
which is a constant, so that \eqref {eq:critical} is not valid any more.
Define
$\nu^* = \frac {\lceil \nu \lambda \rceil}{\lambda}
< \nu + \frac {1}{\lambda}$; then \eqref {eq:critical}
holds with $\nu^*$ in the place of $\nu$.

We now have to optimise the constant in the subexponential function
giving the total complexity, $2b$, subject to \eqref {eq:minsieve}
and \eqref {eq:sieve=linalg}, in which all occurrences of $\nu$ have
been replaced by $\nu^*$. As with $\nu$ we loose one degree of freedom,
the solution to the optimisation problem becomes worse, and we will
end up with a higher total complexity. In fact, the two equations
\eqref {eq:minsieve} and \eqref {eq:sieve=linalg} in two variables $b$
and $\delta$ admit a unique solution $b$, $\delta > 0$, which is easily
computed. The analysis of the individual logarithm computation step is
modified along the same lines, with an increased effort value.

It is interesting to study what happens when $\lambda \to 0$. This
entails
$\nu^* \sim \frac {1}{\lambda} \to \infty$ (here, $\sim$
denotes equivalence in the sense that the quotient of the left and
the right hand side tends to $1$).
The solution to equations~\eqref{eq:minsieve} and~\eqref{eq:sieve=linalg}
is uniquely determined by $\nu^*$ and yields in particular
\[
b \sim \sqrt {\frac {\kappa}{3}} \cdot \frac {1}{\sqrt \lambda}.
\]
Similarly, in the special-$Q$ descent step, we have
$$\deg_Y\varphi =k = \sigma \frac {n}{(g/\M)^{1/3 - \tau/2}}
= \sigma \lambda (g/\M)^{\tau / 2}.$$
Assuming the worst case scenario, which is $\tau$ very close to $0$
(corresponding to the end of the descent), we must ensure that
$\sigma\lambda\geq1$. We thus have to
replace the optimal $\sigma$ by $\sigma^*\sim\frac1\lambda$. This changes
the equation giving $c'$ as a function of $c$. For the limit of the
sequence $c_i$ to match $b$, we thus have to adapt the effort value $e$.
We obtain:
$$e\sim
\sqrt{\frac{\kappa}{3}}\cdot\frac1{\sqrt\lambda}.$$

Given that $b$ and $e$ tend to infinity when $\lambda \to 0$,
we expect that a complexity of $L(1/3)$ will no longer be achievable
using the presented algorithm when $n$ grows more slowly than $(g/\M)^{1/3}$;
this is confirmed by the following analysis.

\subsection{$n$ below $(g/\M)^{1/3}$}
\label {ssec:overcritical}

When the lower bound for $n_i$ has the form $\lambda(g/\M)^\alpha$
with $\alpha<1/3$, then we have $d=\log_qL(1-\alpha,O(1))$ at best. This
implies that in the algorithm depicted in this article, both in the
relation collection and individual logarithm steps, the best possible
upper bound for the norm of the
functions $\varphi$ is $\deg_X\Norm(\varphi)\leq\log_qL(1-\alpha,O(1))$.
We then obtain an algorithm of complexity
$$L\left(\tfrac{1-\alpha}2,c\right)\text{ for some $c>0$.}$$
Following exactly the lines of the proofs of
Theorems~\ref{th:relations} and~\ref{th:logarithms},
it is also possible to make the constant
$c$ in the expression above completely explicit.

\subsection{Curves with a low weighted degree}

\begin {theorem}
Assume that the family of curves of Theorem~\ref{th:relations}
satisfies the following additional constraint: $\kappa = 2$, and each monomial
$X^j Y^k$ occurring in the equation of $\C$ has
$n j + d k \leq nd$.
For instance, the curves may be $\C_{nd}$ curves.

Then the relation collection and the linear
algebra phases are performed in time
$L_{q^g} (1/3, c + o (1))$ with
$c = \sqrt[3] {\frac {64}{9}}$.
\end {theorem}

\begin{remark}
The case of plane non-singular curves of total degree $\approx \sqrt g$,
which has been studied by Diem in~\cite{Diem06ecc}, is included in the
theorem. In this case, one has additionally $n \approx d \approx \sqrt g$
and $\alpha = 1/2$.
\end{remark}
\medskip

\begin {proof}
We use the notation of the proof of Theorem~\ref{th:relations}.
Instead of bounding the degrees of $X$ and $Y$ in $\varphi$
separately (``taking $\varphi$ from a rectangle''), we
take $\varphi$ of bounded weighted degree (``from a triangle'').
The monomials $X^j Y^k$ occurring in $\varphi$ are required to
satisfy $n j + d k \leq \lambda g^{2/3} \M^{1/3}$ for
some parameter $\lambda$ replacing $\nu$ and $\delta$ and to be
optimised later.

Then
\[
\deg_X \Res_Y (\varphi, \C)
\leq \lambda g^{2/3} \M^{1/3}
= \log_q L_{q^g} (2/3, \lambda),
\]
which yields a smoothness probability of
\[
L \left( 1/3, - \frac {\lambda}{3 b } + o(1) \right).
\]

The biggest power of $X$ in $\varphi$ is
$\frac {\lambda g^{2/3} \M^{1/3}}{n}$,
the biggest power of $Y$ is
$\frac {\lambda g^{2/3} \M^{1/3}}{d}$. The number
of allowed monomials is given by the product of these two
quantities divided by $2$,
so that the search space has size about
\[
q^{\frac {\lambda^2 g^{4/3} \M^{2/3}}{2 n d}}
\geq q^{\lambda^2 g^{1/3} \M^{2/3} / (2 \kappa)}
= L (1/3, \lambda^2 / 4).
\]
So the expected number of relations becomes
$L \left( 1/3, \lambda(3b\lambda-4)/12b\right)$,
which should be the same as the factor base size. Thus,
$b = \lambda(3b\lambda-4)/(12b)$.
Equating the time spent in the relation collection
and in the linear algebra phase, we get
$ {\lambda^2}/{4} = 2 b$.
These two equations are solved by
$$
b = \sqrt [3] {\frac {8}{9}} \quad
\text{and}\quad
\lambda  =  \sqrt [3] {\frac {64}{3}}
$$
and yield a total complexity of $L (1/3, c)$ with
\[
c = 2 b
= \sqrt [3] {\frac {64}{9}}.
\]
\end{proof}

To conclude, we note that the runtime for computing individual
logarithms by special-$Q$ descent derived in Section~\ref {sec:logarithms}
is still dominated by the improved runtime for relation collection and
linear algebra in this special case.
Therefore, while an analogously improved approach to individual logarithms
using functions ``from a triangle'' would work, it would not have any
effect on the total complexity, and we omit its analysis.

\paragraph {Acknowledgement.} We thank an anonymous referee for helpful
suggestions.

\end{document}